\documentclass[10pt,journal,cspaper,compsoc]{IEEEtran}
\usepackage[ruled]{algorithm2e}
\usepackage{graphicx}
\usepackage{hyperref}
\usepackage{cite}
\usepackage{breakurl}
\usepackage{graphicx,amssymb,amsmath,lineno}
\usepackage{float}
\usepackage{subfigure}
\usepackage{color}

\ifCLASSOPTIONcompsoc
\else
\fi
\ifCLASSINFOpdf
\else
\fi
\begin{document}
%
\title{Probe Machine}
%
%
%
%

\author{\IEEEauthorblockN{Jin Xu}\\
\IEEEauthorblockA{\IEEEauthorrefmark{1}School of Electronic Engineering and Computer Science, Peking University, Beijing 100871,  China; Key Laboratory of High Confidence Software Technologies (Peking University), Ministry of Education, CHINA}}

\IEEEcompsoctitleabstractindextext{%
\begin{abstract}
A novel computing model, called \emph{Probe Machine}, is proposed in this paper. Different from Turing Machine, Probe Machine is a fully-parallel computing model in the sense that it can simultaneously process multiple pairs of data, rather than sequentially process every pair of linearly-adjacent data.
In this paper, we establish the mathematical model of Probe Machine as a 9-tuple consisting of data library, probe library, data controller, probe controller, probe operation, computing platform, detector, true solution storage, and residue collector. We analyze the computation capability of the Probe Machine model, and in particular we show that Turing Machine is a special case of Probe Machine. We revisit two NP-complete problems---i.e., the Graph Coloring and Hamilton Cycle problems, and devise two algorithms on basis of the established Probe Machine model, which can enumerate all solutions to each of these problems by only one probe operation. Furthermore, we show that Probe Machine can be implemented by leveraging the nano-DNA probe technologies. The computational power of an electronic computer based on Turing Machine is known far more than that of the human brain. A question naturally arises: will a future computer based on Probe Machine outperform the human brain in more ways beyond the computational power?
\end{abstract}

}

\maketitle

\IEEEdisplaynotcompsoctitleabstractindextext

%
\IEEEpeerreviewmaketitle

\section{Introduction}
%
%

%
%
%
%
\IEEEPARstart{T}{he} calculation tool is known as one of the primary factors driving the development of human civilization, and it gradually evolves as the human civilization advances. Human development can be divided into five stages by technological milestones: namely stone age, iron age, steam age, electric age, and information age. Accordingly, the calculation tool  has experienced a long evolutionary process from simplicity to complexity, successively appearing in the forms of knotted rope (for simple reckoning), abacus, slide rule, mechanical computer, and electronic computer, each of which plays an indispensable role in its historical period.

Computer is conceptually a general purpose device that is built upon a computing model, and manufactured by certain materials that can be used to implement the specific computing model. For instance, today's electronic computer is conceptualized by Turing Machine (TM) \cite{Turing} and composed of electronic components \cite{Turing2}.

\begin{figure} [H]
\centering
 \includegraphics[width=200pt]{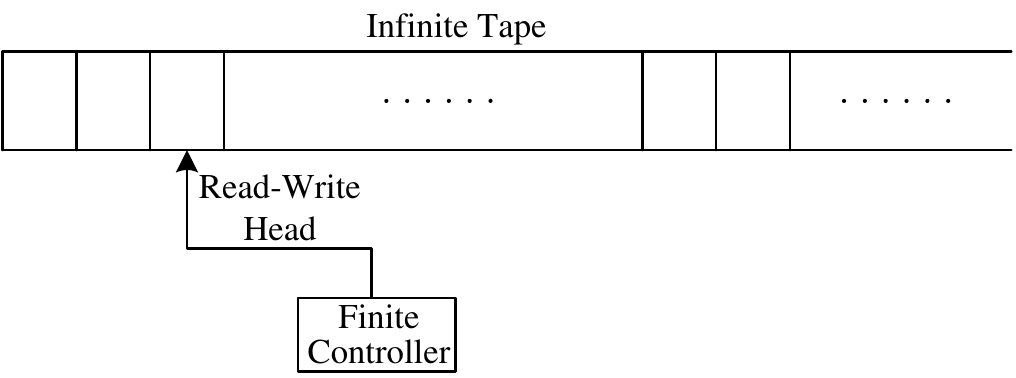}
 \caption{A schematic diagram of Turing Machine.}
  \label{figure1}
\end{figure}

TM was introduced by Alan Turing in order to address the problem of determining whether a given polynomial Diophantine equation with integer coefficients has an integer solution. This problem is among the Hilbert's twenty-three problems in mathematics, which were announced in the International Congress of Mathematicians held in 1900. As shown in Figure \ref{figure1}, TM consists of three components: a finite controller, an infinite tape, and a read-write head.

In 1945, Von Neumann built up the architecture of electronic computer, using TM as its computing model and semiconductors as its implementation materials. In the following year, the first electronic computer based on the Von Neumann architecture was successfully developed. Till now, the development of electronic computer has experienced four main stages, including tube computer, transistor computer, integrated circuit computer, and massive integrated circuit computer.

Although the phenomenal growth of TM-based computing devices (e.g., electronic computer) follows Moore's Law \cite{3}, scientists have been seeking new computing models for building more powerful computers during the previous half century. In 2011 when the 100th anniversary of Turing's birth, an open solicitation was sent to the world calling for new models of computation going beyond TM. The motivations behind the pursuit of new computing models are twofold: (1) the manufacturing technology of semiconductors will reach its limit in next ten years, as predicted by Kaku in 2012; and (2) today's electronic computers have been unable to solve large-scale NP-complete problems due to the limitation of TMs.

In the exploration of new computing models, bionic computing (e.g., neural networks, evolutionary computing, particle swarm optimization and computing), optical computing, quantum computing, and biological computing model have been proposed in literature. Note that bionic and optical computing models are comparable to TM in terms of their computation capability~\cite{4,5,6}. Given an algorithm with a computational complexity of $n$ under TM, the quantum computing model is capable of reducing the computational complexity to $\sqrt{n}$ \cite{7,8,9}; this implies that the quantum computing model does not surpass TM with respect to computability. The biological computing model provides an opportunity of solving large-scale NP-complete problems within a short time, owing to its nano-DNA data cell and the significant parallelism of computation given by DNA specific hybridization. Many DNA computing models can significantly boost the computation speed by making use of nucleic acid molecules, including Adleman's model \cite{10}, sticky model \cite{11}, self-assembly model \cite{12,13,14}, non-enumerated model \cite{15} and parallel DNA model \cite{16}; however, there exist many non-solutions in the output of these DNA computing models. It has been shown that the aforementioned computing models are equivalent to TM with respect to computability, while they lead to different levels of computational efficiency.

In this paper, we propose a novel computing model, called \emph{Probe Machine (PM)}, which is fully parallel as it can process an arbitrary number of pairs of data simultaneously. We identify two features of TM that limit its computability. Then we establish the mathematical model of PM that overcomes these two problems, analyze the performance of PM, and show that TM is a special case of PM. Finally, we exemplify that Probe Machine can be implemented by using nano-DNA probe technologies; meanwhile, it can easily solve two NP-complete problems at a very small computational complexity.

\section{Featured Mechanisms of TM}

It is challenging to answer the following two questions: (1) Why does a TM-based computer generate a large number of non-solutions when solving NP-complete problems? (2) Why are the existing computing models equivalent to the TM model? Next, we identify two features of TM mechanisms that explain the limited computability of TM-based computers, and the equivalence among existing models of computation.

\subsection{Linear Data Placement Mode}

Any existing computing device depends on a conventional way of storing data, which is called the \emph{data placement mode}, during its computation process.

In the development of human civilization over thousands of years, words, letter, or numbers are typically written from left to right, or from top to bottom, or one next to another. In other words, data units are placed or stored linearly. Possibly influenced by such a ``conventional mindset", the data used by human's calculation tools are similarly placed one next to another, no matter in which form the tool appears: knotted ropes, stones for simple reckoning, abacus, or today's electronic computer. We call such a data placement mode as \emph{linear}.

In this linear data placement mode, only adjacently-placed data can be processed simultaneously, which greatly limits the computation capabilities of calculation tools. All of existing computing models suffer from this limitation. For example, when a computer program handles a NP-complete problem, a substantial number of feasible solutions it outputs in the (initial) solution space are not optimal, and searching for true solutions is like looking for a needle in a haystack.

Therefore, the linear data placement mode is the root cause of the difficulty in seeking true solutions, and the generation of vast non-solutions in the (initial) solution space.

\subsection{Sequential Data Processing Mode}

Given the linear data placement mode, a calculation tool under TM can only process adjacently-placed data in a single operation. We call this type of data processing mode as \emph{sequential}. The sequential data processing mode has become another root cause of the generation of vast non-solutions in the (initial) solution space.

Since the data placement and data processing modes in the bionic, quantum, and DNA computing models mentioned above are the same as those employed by TM, all these computing models are essentially equivalent to the TM model.

To sum up, TM has two featured mechanisms: (1) the linear data placement mode, and (2) the sequential data processing mode, which in turn imposes constraints on the computation capability of the TM model.

Hence, it is necessary to break these two constraints in search of a computing model that is fundamentally more powerful and effective than TM. Accordingly, there is a central design requirement for devising such a conceptually brand-new model---the model needs to be capable of simultaneously processing as much adjacently-placed data as possible. That is, the way of placing data should be non-linear, which is the main motivation of proposing Probe Machine in this paper.

\section{The Mathematical Model of Probe Machine}


We define a Probe Machine ($PM$) as a 9-tuple:
$$PM=(X, Y, \sigma_1, \sigma_2,\tau, \lambda, \eta, Q, C),$$
where the 9 components denote the data library ($X$), probe library ($Y$), data controller ($\sigma_1$), probe controller ($\sigma_2$), probe operation ($\tau$), computing platform ($\lambda$), detector ($\eta$), true solution storage ($Q$), and residue collector ($C$), respectively. In the following, we will explain each component of PM in details. 

\subsection{Data library}

\begin{figure} [H]
\centering
 \includegraphics[width=240pt]{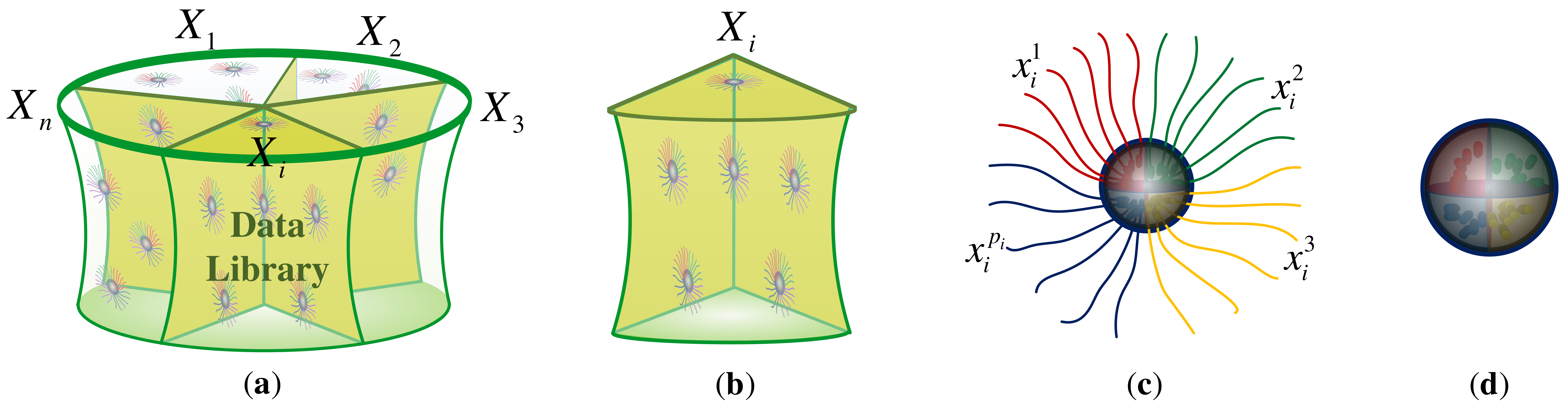}
 \caption{(a) Data library. (b)  Data pool $X_i$. (c) Data $x_i$ that has a number of data fibers. (d) Data body. }
  \label{figure2}
\end{figure}

The data library of PM defines a new data placement mode that is non-linear. The data library $X$ consists of $n$ data pools $X_1, X_2, \cdots, X_n$, that is, $X=\{X_1, X_2, \cdots, X_n\}$ (see Figure \ref{figure2} (a)).
Each data pool $X_i$ contains one type of data $x_i$ (see Figure \ref{figure2} (b)). When we only consider the types of date of $X$, $X$ is viewed as a set of $n$ elements, denoted by $X=\{x_1, x_2, \cdots, x_n\}$, and $X_i$ is a very large set.

Each $x_i$ contains a \emph{body} and numerous data \emph{fibers}. There are $p_i$ types of data fibers, denoted by $x_i^{1},x_i^{2},\cdots,x_i^{p_i}$, respectively. Each $x_i$ contains massive copies of each type of data fiber.
The data body is divided into several regions, and each region only contains the same type of data fibers. Figure \ref{figure2} (c) shows the structure of data $x_i$, where the ``sphere'' represents the data body of $x_i$ (see Figure \ref{figure2} (d)) and ``colored lines'' represent different types of data fibers.
Let $\Im(x_i)=\Im_i$ denote the set of $p_i$ types of data fibers of $x_i$, namely, $\Im(x_i)=\{x_i^{1},x_i^{2},\cdots,x_i^{p_i}\}$.

Then we define data $x_i$ as
\begin{equation}
x_i=(i,x_i^1,x_i^2,\cdots, x_i^{p_i})=(i; \mathfrak{J}_i)
\end{equation}
for $i=1,2,\cdots, n$, where $i$ is the body of $x_i$. Let $p$ be the number of different types of data in the data library. It then follows that

\begin{equation}
p=p_1+p_2+\cdots+p_n
\end{equation}

Each data library $X$ in PM is associated with the following four properties.
\begin{enumerate}
  \item[(1)] $X$ contains $n$ different types of data and $p$ different types of data fibers;
  \item[(2)] Each data pool $X_i$ contains a large number of data $x_i$ for $i=1,2,\cdots, n$;
  \item[(3)] Each data pool contains a controllable output system, called \emph{data controller};
  \item[(4)] Data library $X$ can be written in the form of matrix as follows,
\end{enumerate}

\[\begin{array}{l}
X=\left(
  \begin{array}{cccc}
    x_1^1 & x_2^1 & \cdots & x_n^1 \\
    \vdots & \vdots & \ddots & \vdots \\
    x_1^{p_1} & x_2^{p_2} & \cdots & x_n^{p_n} \\
  \end{array}
\right)
\end{array}\\
\]
where the $i$th (1$\leq i\leq n$) column contains the $p_i$ types of data fibers belonging to $x_i$. It is possible that $p_i$'s are different, and thus $X$ has $n$ columns and variable number of rows.

\subsection{Probe library}

Generally speaking, the ``probe'' is a tool used to detect certain substances. The concept of probe has been applied in a wide variety of fields in science, such as biology, computer science, electronics, information security, archaeology, etc. Here we only explain the probe concept in two fields of biology and electronics.

The \emph{probe in biology} is a short \emph{single-stranded DNA} or \emph{RNA} fragment (about 20bp to 500bp), which is designated to detect its complementary DNA sequence. After being heated, a double-stranded DNA denatures and becomes a short-stranded DNA, or synthesizes a single-stranded DNA molecule. The denatured DNA is marked with radioisotope, fluorescent dyes, or enzyme as a probe. The biological probe detection process is as follows: first, the probe  hybridizes with samples, which leads to the probe and its complementary DNA sequence are closely linked through hydrogen bonds; second, the redundant probe not being hybridized is washed away; finally, depending on the types of probes, the detection methods, such as autoradiography, fluorescence microscope and enzyme-linked amplification, can be used to detect whether the sample contains the test sequence or which positions contain the test sequence (namely, the sequence that is complementary to the probe).

The \emph{probe in electronics} can be classified into the following three different categories according to the purposes of electronic test:
\begin{enumerate}
\item[(1)] Light circuit board test probe is used to test circuit board or detect short circuits;
\item[(2)] Online test probe is used to test components installed on PCB circuit board;
\item[(3)] Microelectronic test probe is used to test wafer or IC chips.
\end{enumerate}

The \emph{probe in PM} is defined as follows. Let $x_{i}^{l}$ and $x_{t}^{m}$ be two types of data fibers. The \emph{probe} between $x_{i}^{l}$ and $x_{t}^{m}$, denoted as $\tau ^{x_{i}^{l}x_{t}^{m}}$, is an operator  that meets all the following three conditions:

\begin{enumerate}
\item[(1)] The probe $\tau ^{x_{i}^{l}x_{t}^{m}}$ can accurately locate $x_{i}^{l}$ and  $x_{t}^{m}$ in the computing platform $\lambda$ (defined later). This condition is referred to as  \emph{adjacency}.
\item[(2)]  The probe $\tau ^{x_{i}^{l}x_{t}^{m}}$ can (only) find $x_{i}^{l}$ and $x_{t}^{m}$ in the computing platform $\lambda$. This condition is referred to as \emph{uniqueness}.
\item[(3)] The operator (i.e., the probe $\tau ^{x_{i}^{l}x_{t}^{m}}$) can implement some operations (e.g.,  connective or transitive operations) upon finding $x_{i}^{l}$ and $x_{t}^{m}$ in the computing platform. The operation result is denoted as $\tau^{x_{i}^{l}x_{t}^{m}}(x_{i},x_{t})$. This condition is referred to as \emph{potential probe}.
\end{enumerate}


If there exists a probe $\tau^{x_{i}^{\ell}x_{t}^{m}}$ between data fibers $x_{i}^{\ell}$ and $x_{t}^{m}$ in the data library, then $x_{i}^{\ell}$ and $x_{t}^{m}$ are called \emph{potential}, and data $x_{i}$ and $x_{t}$ are called \emph{potential}. Otherwise, $x_{i}^{\ell}$ and $x_{t}^{m}$ are called \emph{nonpotential}.

If for any $\ell$ and $m$, there is no probe among any pair of data fibers $x_{i}^{\ell}$ and $x_{t}^{m}$, then data $x_{i}$ and $x_{t}$ are \emph{nonpotential}.

For any $X'\subseteq X$, we denote by $\tau(X')$ the set of probes for all possible pairs of fibers belonging to data in $X'$, and $\tau(X')$ is called the \emph{probe set} of $X'$.

Then, we define two types of probes of PM: connective and transitive probes.

\begin{figure} [H]
\includegraphics[width=240pt]{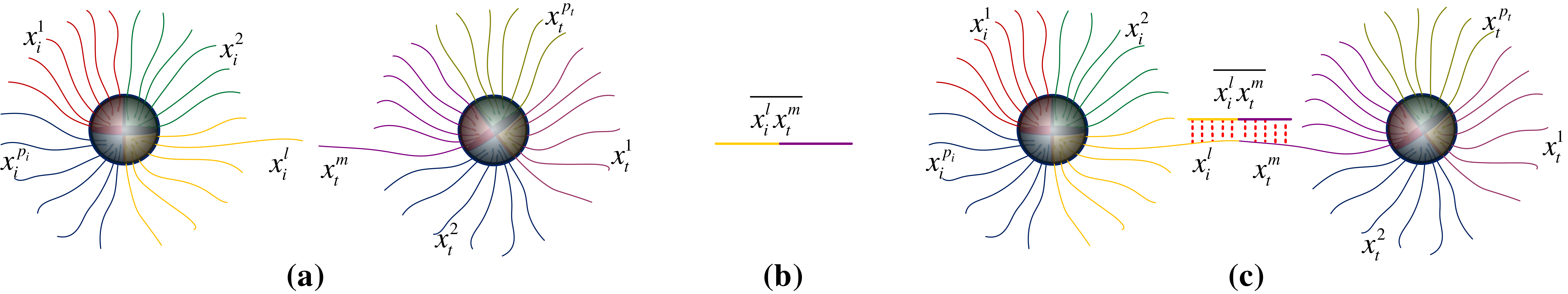}
 \caption{Illustration of connective operator.}
  (a) Data $x_{i}$ and $x_{t}$. (b) Connective probe of $x_{i}^{l}$ and $x_{t}^{m}$. (c) Result of $\overline{x_i^lx_t^m}$.
  \label{figure3}
\end{figure}

A \emph{connective probe}, denoted as $\overline{x_{i}^{\ell}x_{t}^{m}}$, refers to a probe $\tau^{x_{i}^{\ell}x_{t}^{m}}$ \emph{connecting} two target data fibers $x_{i}^{\ell}$ and $x_{t}^{m}$ in the computing platform.
This definition is illustrated in Figure \ref{figure3}, where Figure \ref{figure3}(a) shows the data $x_{i}^{\ell}$ and $x_{t}^{m}$, and
Figure \ref{figure3}(b) shows the connective probe between $x_{i}^{\ell}$ and $x_{t}^{m}$. The connective probe contains two parts: the complement of data fiber $x_{i}^{\ell}$, and the complement of data fiber $x_{t}^{m}$. Since the complement of a data fiber $x_{i}^{\ell}$ has the capability of adsorbing $x_{i}^{\ell}$ itself, two complements forming the connective probe can easily adsorb the two data fibers in $x_{i}$ and $x_{t}$, i.e., it can effectively connect $x_{i}$ and $x_{t}$ through two data fibers. The result of a connective probe is denoted as $\overline{x_{i}^{\ell}x_{t}^{m}}(x_{i},x_{t})\triangleq x_{i}x_{t}^{\overline{x_{i}^{\ell}x_{t}^{m}}}$ (see Figure \ref{figure3}(c)).

Data fibers that can be connected by connective probes are referred to as \emph{connective data fibers}, and a data containing only connective fibers is called a \emph{connective data}. If all data in the data library is connective, the data library is called \emph{connective data library}.

The data fibers $x_i^{\ell}$ and $x_t^m$ shown in Figure \ref{figure3} can be viewed as two different DNA strands embedded in nanoparticle, where data bodies of $x_i$ and $x_t$ can be considered as nanoparticles. The implementation of a connective probe $\overline{x_i^{\ell}x_t^m}$ can be regarded as the hydrogen bonding force among $x_i$, $x_m$ and their complements.

In fact, a connective probe is a formal definition which is abstracted from many instances in the real world. For example, the biological probe mentioned above is an example connective probe. According to the definition, a connective probe does not consider the direction of information processing between data fibers. It only finds two target data fibers and then connects them together.

\begin{figure}[H]
\includegraphics[width=240pt]{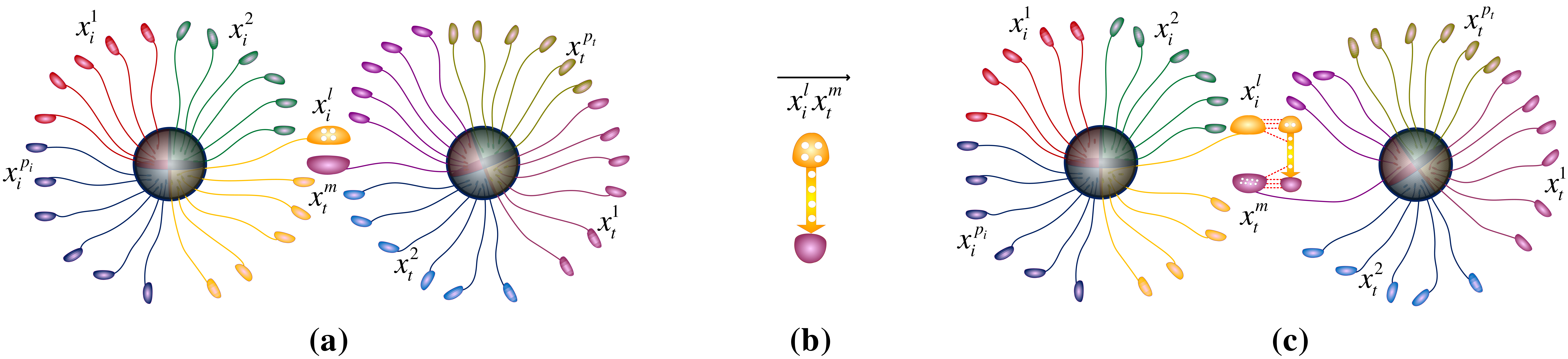}
 \caption{Illustration of transitive probe}
 (a) Data $x_{i}$ and $x_{t}$.
 (b) Transitive probe of $x_{i}^{l}$ and $x_{t}^{m}$.
 (c) Result of $\overrightarrow{x_i^lx_t^m}$.
  \label{figure4}
\end{figure}

We introduce the concept of a transitive probe to characterize the direction of information processing between data fibers. Assume that all data fibers of $x_i$ and $x_t$ carry information, as shown in Figure \ref{figure4}(a).
A probe
${\tau ^{x_i^{\ell}x_t^m}}$ on $x_i^{\ell}$ and $x_t^m$, denoted by $\overrightarrow{x_i^{\ell}x_t^m}$, is called a \emph{transitive probe}, if ${\tau ^{x_i^{\ell}x_t^m}}$ finds two target data fibers $x_i^{\ell}$ and $x_t^m$ in the computing platform  and transmits information from $x_i^{\ell}$ to $x_t^m$, where $x_i^{\ell}$ is
defined as \emph{source fiber} and $x_t^m$ is defined as \emph{destination fiber}, as shown in Figure \ref{figure4}. More specifically,
Figure \ref{figure4}(a) shows the data $x_{i}^{\ell}$ and $x_{t}^{m}$,
Figure \ref{figure4}(b) shows a transitive probe between $x_{i}^{\ell}$ and $x_{t}^{m}$, and
Figure \ref{figure4}(c) represents the result of the transitive probe, denoted by
$\overrightarrow {x_i^{\ell}x_t^m} ({x_i},{x_t}) \buildrel \Delta \over = {x_i}{x_t}\overrightarrow {^{x_i^{\ell}x_t^m}}$.

Data fibers that can be processed by transitive probes are referred to as \emph{transitive data fibers}, and furthermore a data containing only transitive fibers is called a \emph{transitive data}. If all data in the data library is transitive, the data library is called a \emph{transitive data library}.

A probe library is built upon the data library in PM, according to the following principles.

\emph{Principle 1}. For any $i$, any pair of data fibers belonging to the same data $x_{i}$ has no probes.

\emph{Principle 2}. For any $i$ and $t$ with $1\leq i, t\leq n$ and $i\neq t$, any pair of data fibers $x_{i}^{l}$ and $x_{t}^{m}$ of different data $x_{i}$ and $x_{t}$ are possibly potential.

A probe library based on a connective data library is called a \emph{connective probe library}, and a probe library based on a transitive data library is called a \emph{transitive probe library}.

It is easy to show that a connective probe library contains at most $|E(K_{p_{1},p_{2},\cdots,p_{n}})|=\frac{1}{2} \{p^2 -(p_{1}^{2}+p_{2}^{2}+\cdots+p_{n}^{2})\}$ probes; and a transitive probe library contains at most $|A(\overrightarrow{K_{p_{1},p_{2},\cdots,p_{n}}})|=p^2 -(p_{1}^{2}+p_{2}^{2}+\cdots+p_{n}^{2})$ probes, where $E(G)$ stands for the edge set of a graph $G$, $A(D)$ stands for the arc set of a directed graph $D$, and $K_{p_{1},p_{2},\cdots,p_{n}}$ (or $\overrightarrow{K_{p_{1},p_{2},\cdots,p_{n}}}$) stands for the complete $n$-partite (directed) graph.

The set of all possible probes between two data $x_i$ and $x_t$ in a probe library $Y$ is defined as the \emph{$(i,t)$-complete probe sub-library}, denoted as $Y_{it}$, where $i,t=1,2,\cdots,n$ and $i\neq t$.
For simplicity, we call $Y_{it}$ a \emph{complete probe sub-library} of $Y$.

\begin{figure}[H]
\centering
\includegraphics[width=220pt]{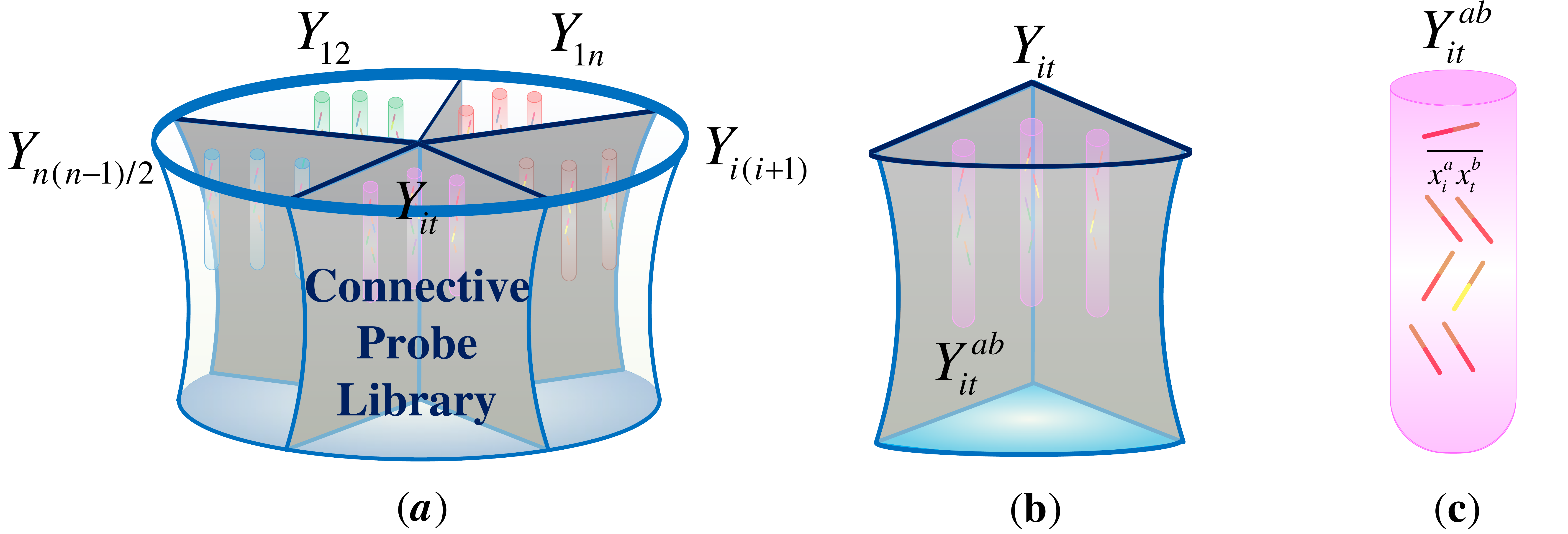}
\caption{Illustration of a connective probe library.}
(a) A connective probe library.
(b) A complete probe sub-library $Y_{it}$.
 (c) A probe pool $Y_{it}^{ab}$ in $Y_{it}$.
\label{figure5}
\end{figure}

As for connective probe libraries, $Y_{it}=Y_{ti}$ holds for any two probe sub-libraries, $Y_{it}$ and $Y_{ti}$. Therefore, there are a total of  $\frac{1}{2}n(n-1)$ complete probe sub-libraries, as shown in Figure \ref{figure5}(a). Figure \ref{figure5}(b) shows $Y_{it}$, which has a total of $\mid Y_{it}\mid=p_{i}\times p_{t}$ types of probes, where

\noindent $
Y_{it}= \Big{\{} \overline{x_{i}^{1}x_{t}^{1}},\overline{x_{i}^{1}x_{t}^{2}},\cdots,\overline{x_{i}^{1}x_{t}^{p_t}};\overline{x_{i}^{2}x_{t}^{1}},
\overline{x_{i}^{2}x_{t}^{2}},\cdots, \overline{x_{i}^{2}x_{t}^{p_t}};$
\begin{equation}
\cdots; \overline{x_{i}^{p_i}x_{t}^{1}},\overline{x_{i}^{p_i}x_{t}^{2}},\cdots,\overline{x_{i}^{p_i}x_{t}^{p_t}} \Big{\}}
\end{equation}

For each type of probes $\overline{x_{i}^{a}x_{t}^{b}}$, we establish a
probe pool $Y_{it}^{ab}$, which contains a large amount of probes $\overline{x_{i}^{a}x_{t}^{b}}$, for $a=1,2,\cdots,p_i$ and $b=1,2,\cdots,p_t$, as shown in Figure \ref{figure5}(c).

\begin{figure}[H]
\includegraphics[width=220pt]{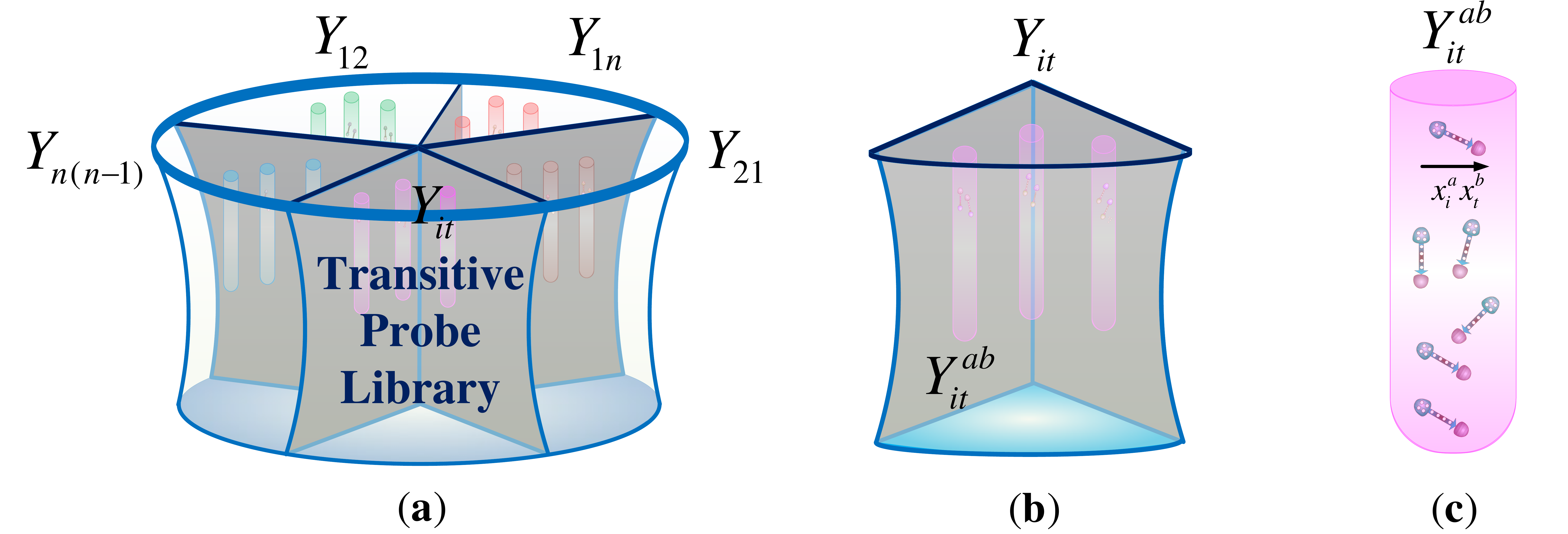}
\caption{Illustration of a transitive probe library.}
(a) A transitive probe library contains $n(n-1)$ complete probe sub-libraries.
(b) A complete probe sub-library $Y_{it}$.
 (c) A probe pool $Y_{it}^{ab}$ in $Y_{it}$.
\label{figure6}
\end{figure}

As for transitive probe libraries, $Y_{it}\neq Y_{ti}$ holds. Therefore, there are a total of $n(n-1)$ complete probe sub-libraries, as shown in Figure \ref{figure6}(a).
 Figure \ref{figure6}(b) shows $Y_{it}$. The probe type number of complete probe sub-libraries in a transitive probe library is exactly twice of that of connective probe libraries. In other words, $Y_{it}$ has a total of $\mid Y_{it}\mid=2p_{i}\times p_{t}$ types of probes, where

\noindent $Y_{it}=\Big{\{}\overrightarrow{x_{i}^{1}x_{t}^{1}},\overrightarrow{x_{t}^{1}x_{i}^{1}},\cdots,\overrightarrow{x_{i}^{1}x_{t}^{p_{t}}},
\overrightarrow{x_{t}^{p_{t}}x_{i}^{1}},\cdots,\overrightarrow{x_{i}^{p_{i}}x_{t}^{1}},\overrightarrow{x_{t}^{1}x_{i}^{p_{i}}},$
\begin{equation}
\cdots, \overrightarrow{x_{i}^{p_{i}}x_{t}^{p_{t}}},\overrightarrow{x_{t}^{p_{t}}x_{i}^{p_{i}}}\Big{\}}
\end{equation}

For each type of probes $\overrightarrow{x_{i}^{a}x_{t}^{b}}$, we establish a
probe pool $Y_{it}^{ab}$, which contains only massive amounts of probes $\overrightarrow{x_{i}^{a}x_{t}^{b}}$, for $a=1,2,\cdots,p_i$ and $b=1,2,\cdots,p_t$, as shown in Figure \ref{figure6}(c).

\subsection{Data controller and probe controller}

\emph{Data controller}, denoted by $\sigma_{1}$, refers to a controller that takes data from data pool and adds it into the computing platform. Each data pool contains one controller, and thus the number of controllers in the data library is equal to the number of data pools---a total of $n$ data controllers.

\emph{Probe controller}, denoted by $\sigma_{2}$, refers to a controller that takes probes from probe pool and adds it into the computing platform.
Each probe pool contains one controller. Therefore, the number of probe controllers in a probe library is equal  to the number of probe pools. A connective probe library contains a total of $1/2\{p^{2}-(p_{1}^{2}+p_{2}^{2}+\cdots+p_{n}^{2})\}$ probe controllers, while a transitive probe library contains a total of  $p^{2}-(p_{1}^{2}+p_{2}^{2}+\cdots+p_{n}^{2})$ probe controllers.


\subsection{Probe Operation}

For a pair of potential data  $x_i$  and $x_t$,  let $x_i^{\ell}$ and  $x_t^m$ be data fibers of $x_i$ and $x_t$, respectively.
\emph{The basic probe operation} consists of the preparing stage and the executing stage.
\begin{itemize}
  \item In the preparing stage, $x_i$, $x_t$, and probe $\tau^{x_i^{\ell}x_t^m}$ are added into computing platform by the data and probe controllers, where $\tau^{x_i^{\ell}x_t^m}$ is responsible for finding $x_i^{\ell}$ and  $x_t^m$.
  \item Once the preparing stage is completed, the executing stage starts, during which the probe is implemented.
\end{itemize}

Generally speaking, \emph{a probe operation} is a process of executing many basic probe operations simultaneously. The formal definition of \emph{a probe operation} is given below. Let $X^\prime$ and $Y^\prime$ be the subset of $X$ and $\tau(X)$, respectively.
A \emph{probe operation} on $X^\prime$ and $Y^\prime$, denoted by $\tau$, is the process of implementing basic probe operations simultaneously, where each basic operation is executed by a probe in $Y^\prime$ on a pair of data in $X^\prime$. The result of the probe operation $\tau$ is called the solution of $\tau$, denoted by $\Theta$,
\begin{equation}
\tau (X',Y') = \Theta
\end{equation}

Note that $\Theta$ is closely related to the computing platform. Indeed, the probe operation can be considered as a \emph{reaction process}: (1) the object is the data-set $X'$; (2) the executors are the whole of probes in $Y'$; and (3) the carrier for realizing this reaction is a computing platform that will be introduced next.

\subsection{Computing platform}

\begin{figure}[H]
\centering
\includegraphics[width=195pt]{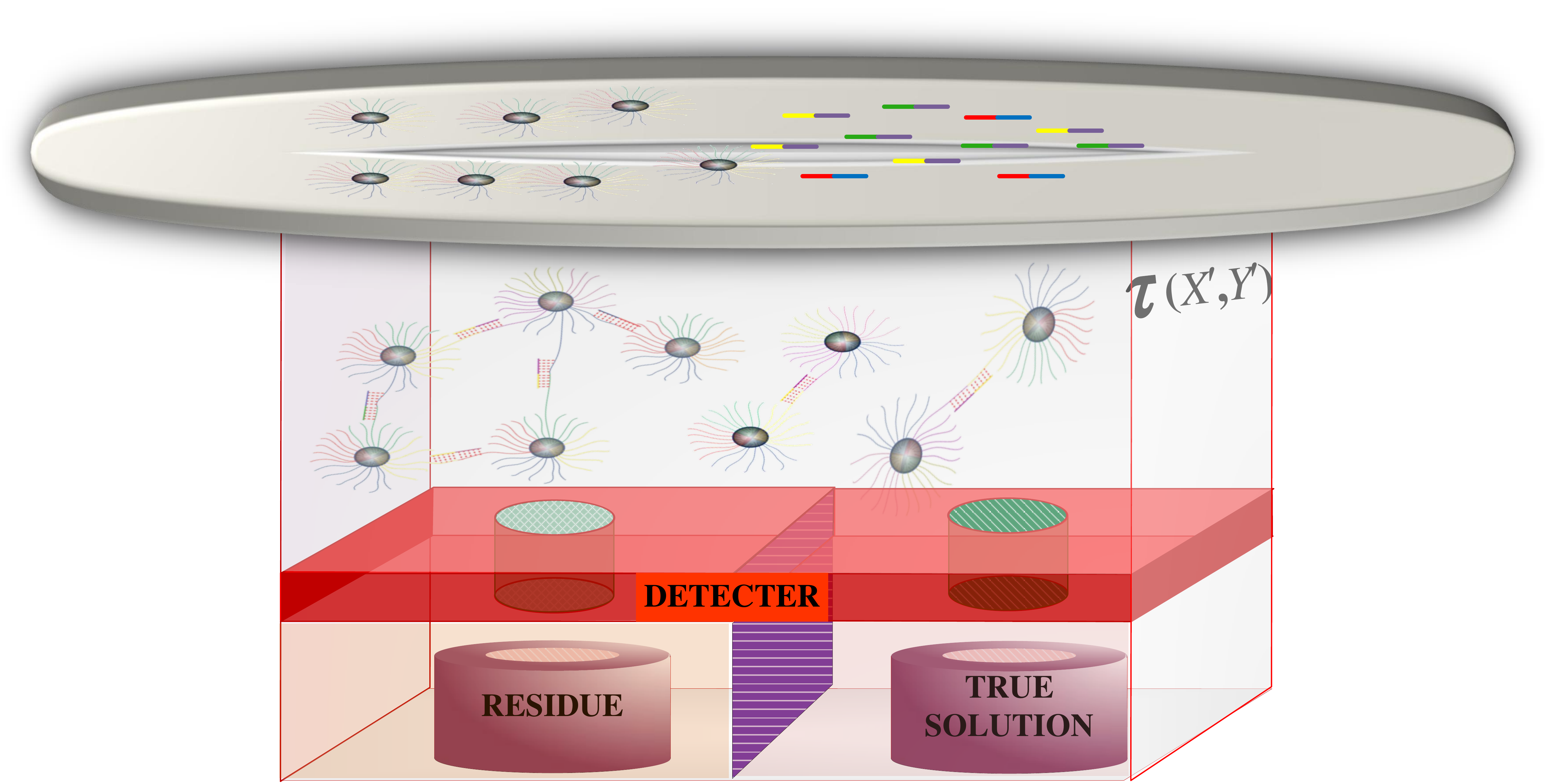}\\
\textbf{(a)}\\
\includegraphics[width=195pt]{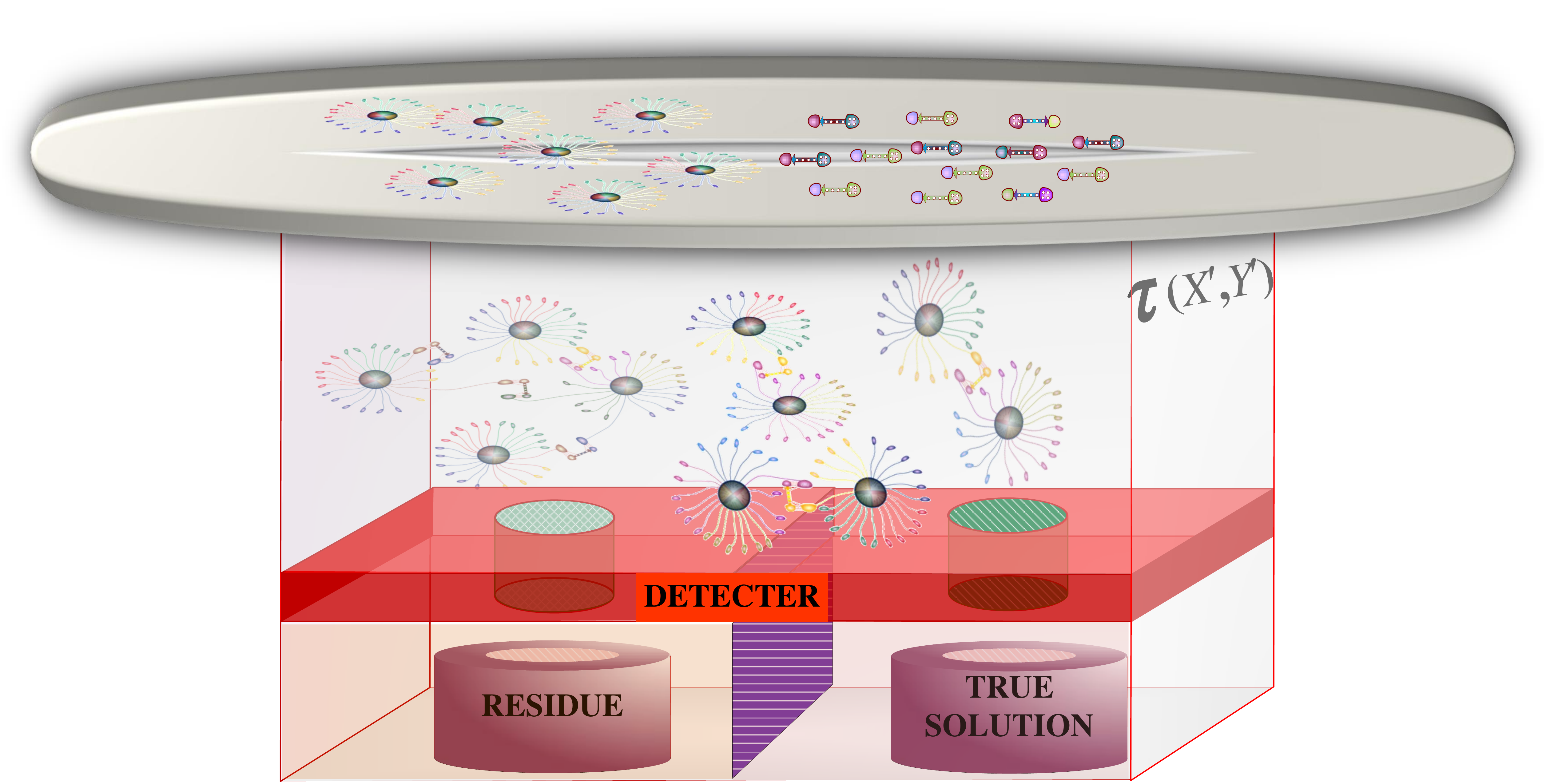}\\
\textbf{(b)}
\caption{Illustrations of computing platform, detector, true solution storage, and residue collector.}
\label{figure7}
\end{figure}

The \emph{computing platform}, denoted by $\lambda$, is an environment that is designated for conducting probe operations. It can assist probes to find target data fibers rapidly and then conduct (basic) probe operations (see Figure \ref{figure7}).

We refer to the aggregation consisting of the two potential data and corresponding probes as a \emph{2-aggregation}. Similarly, the aggregation consisting of a 2-aggregation and a third data, together with corresponding probes, is called a \emph{3-aggregation}; in particular, each data is called a \emph{$1$-aggregation}. An \emph{$m$-aggregation} contains $m$ data, and corresponding probes that find the $m$ data. For any $m$-aggregation $M$, we call $m$ the \emph{order} of $M$, denoted by $|M|$, i.e. $|M|=m$.

The computing platform has the following three fundamental functions.

Fundamental Function 1: High Cohesiveness. When a probe ${\tau ^{x_i^lx_t^m}}$ is added into the computing platform, this probe always searches for
 two target data fiber $x_i^l$ and $x_t^m$, which can produce high-order multi-probe aggregation. Specifically, there are two rules to be followed.

(I) Let ${M_1},{M_2},{M_3}$  be three aggregations in the computing platform. Suppose that $x_i^l$ is a data fiber of $M_1$,
   and $x_t^m$ is a data fiber of $M_2$ and $M_3$ as well. Without loss of generality, we assume $|M_2|> |M_3|$. 
 Then  ${\tau ^{x_i^lx_t^m}}$ will choose $x_i^l$ from $M_1$ and $x_t^m$ from $M_2$, to conduct the basic probe operation.

(II) If both $M_1$ and $M_2$ contain data $x_i$ and $x_t$, and the probe  ${\tau ^{x_i^lx_t^m}}$ can only conduct the basic probe operation on $M_1$ and $M_2$:
 \begin{enumerate}
   \item[(1)] When $|M_1|>|M_2|$, ${\tau ^{x_i^lx_t^m}}$ will choose two data $x_i$ and $x_t$ from $M_1$ to conduct the basic probe operation.
   \item[(2)] When $|M_1|=|M_2|$, ${\tau ^{x_i^lx_t^m}}$ will choose two data $x_i$ and $x_t$ from the
    aggregation that has more probes than the other, to conduct basic probe operation.
   \item[(3)] When $|M_1|=|M_2|$ and they contain the same number of probes, ${\tau ^{x_i^lx_t^m}}$ will
   choose data randomly from either of the two aggregations to conduct basic probe operation.
 \end{enumerate}
The size of an aggregation has a limit, which is restricted by the following threshold property of the computing platform.

Fundamental Function 2: Threshold Property. Each aggregation's order in the computing platform needs to be no greater than  $|V(G^{(X',Y')})|$, where $G^{(X',Y')}$ is the probe operation graph (see the subsection of Detector for details).

Let $M_1$ and $M_2$ be two aggregations. If $|M_1|+|M_2|>|V(G^{(X',Y')})|$, the number of data in the two aggregations is greater than the number of data in a true solution, and any basic probe operation on two data in $M_1$ and $M_2$ will not lead to a true solution. In this case, PM cannot allow any basic probe operation on two data in $M_1$ and $M_2$.

Fundamental Function 3: Uniqueness. Given an aggregation $M$ where $|M|\geq 2$ in the computing platform, there is at most one data for each type that $M$ contains; and there exists at most one probe operation between any two data in $M$.

\subsection{Detector}

We define a tool called \emph{the probe operation graph} for detecting true solutions from all generated solutions. A probe operation graph of data subset $X^\prime$ and probe subset $Y^\prime$ , denoted by $G^{(X^\prime,Y^\prime)}$, is the topological structure of the aggregation of true solutions after probe operations. The vertex set $V(G^{(X^\prime,Y^\prime)})$ is the set of data in an aggregation of a true solution and the edge set $E(G^{(X^\prime,Y^\prime)})$ is the set of probes.
The aggregation that is not isomorphic to $G^{(X^\prime,Y^\prime)}$ is called \emph{residues}.

For a given problem, its probe operation graph can be deterministic.
For example, in the solution to  Hamilton cycle problem in Figure \ref{figure9}(c), the probe operation graph is either a 4-cycle or 5-cycle. In the solution to coloring problem of graph $G$, the probe operation graph is just the graph $G$ itself.
In the solution of finding the maximum clique of a given graph, the probe operation graph is the topological structure of the aggregation with maximum order.

A solution is true if and only if its topology is isomorphic to $G^{(X',Y')}$. Therefore, a detector $\eta$ detects all solutions isomorphic to $G^{(X',Y')}$ in computing platform and separates them from residues (see Figure \ref{figure7}). The fundamental functions of a detector $\eta$ are as follows.
 \begin{enumerate}
 \item [(1)]  For an aggregation $M$, if $|M|\neq |V(G^{(X',Y')})|$ or the number of probes in $M$ is not equal to the number of edges in $G^{(X',Y')}$, then the detector separates it into residue collector.
   \item [(2)] If the order of an aggregation is equal  to the number of edges of $G^{(X',Y')}$  and the probe number in the aggregation is equal  to the number of edges of $G^{(X',Y')}$, then the detector separates it into true solution storage.
 \end{enumerate}

\subsection{Procedure of probe operations}

There are four steps in the procedure of probe operations.

\emph{Step} 1. Determine the probe operation graph $G^{(X',Y')}$.
Since the computing platform has three properties of high cohesiveness, threshold and uniqueness, the probability of true solutions being produced in the probe operations is very high.

\emph{Step} 2. Determine the needed number of data of each type in data-set $X'$, and the needed number of probes in probe sub-library $Y'$, respectively, on basis of the probability of producing $G^{(X',Y')}$.

\emph{Step} 3. Use data controller $\sigma_1$ and probe controller $\sigma_2$ to fetch the determined number of data of $X'$ from data pools, and the determined number of probes of $Y'$ from probe pools, and add them into the computing platform $\lambda$.

\emph{Step} 4. Implement the probe operation $\tau(X',Y')$ in the computing platform $\lambda$. The detector $\mu$ screens out true solutions that are isomorphic to $G^{(X',Y')}$, and put them into true solution storage; meanwhile it separates residues into residue collector.

Note that each true solution has the same topology as $G^{(X',Y')}$, while different true solutions may have different weights (i.e., different types of data fibers). All of the aggregations with the topologies isomorphic to $G^{(X',Y')}$ are the solutions to the problem.

\subsection{True solution storage and residue collection}

True solution storage is needed for storing true solutions and outputting them correctly. The residues produced in the probe operations will be collected by \emph{residue collector}, as shown in Figure \ref{figure7}. The residue collector mainly recycles residues, decomposes them back into the form of basic data, and then returns these basic data to relative data pools.

\subsection{Two Architectures of Probe Machine}

\begin{figure}[H]
  \centering
  \includegraphics[width=195pt]{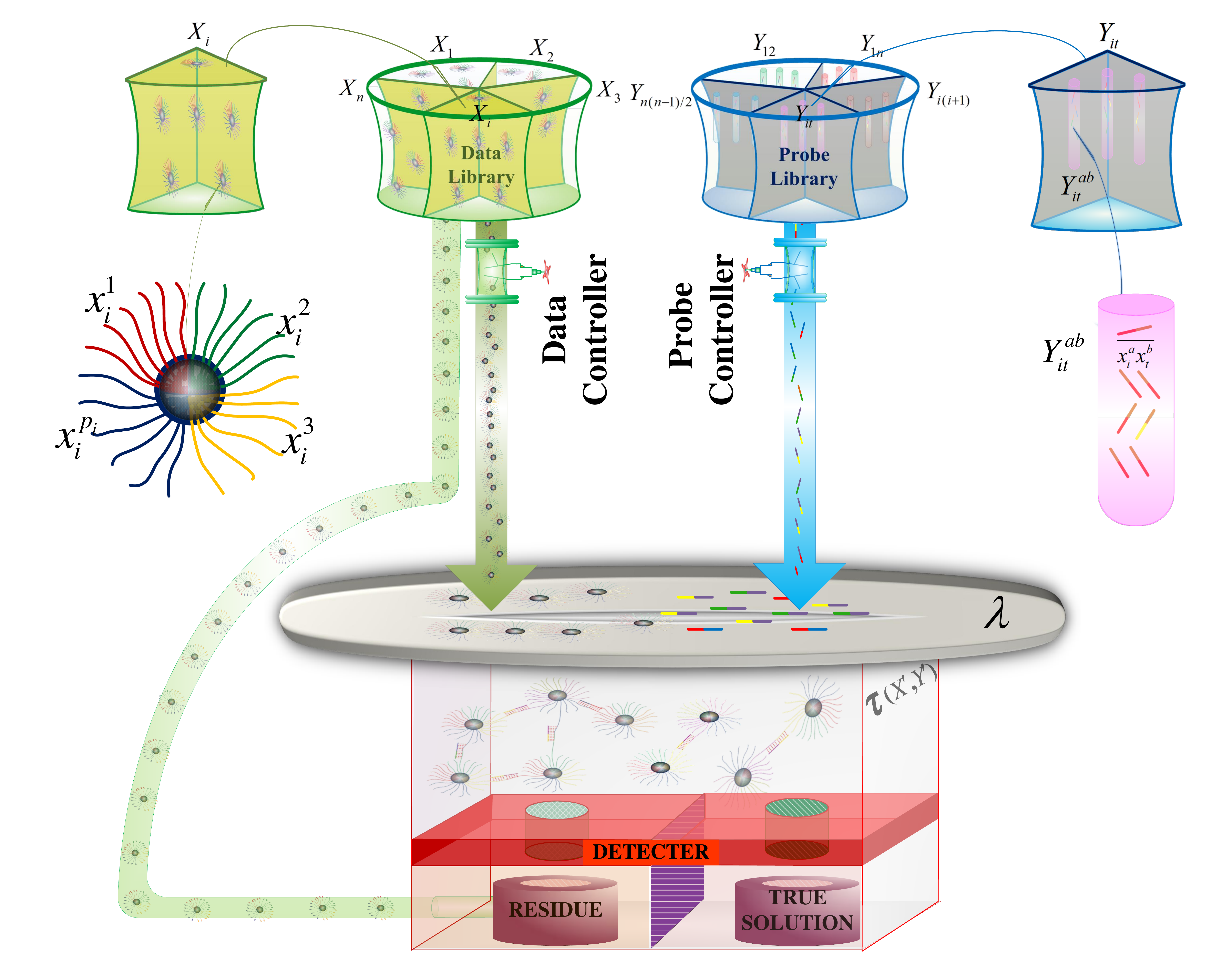}\\
  (a)\\
  \includegraphics[width=195pt]{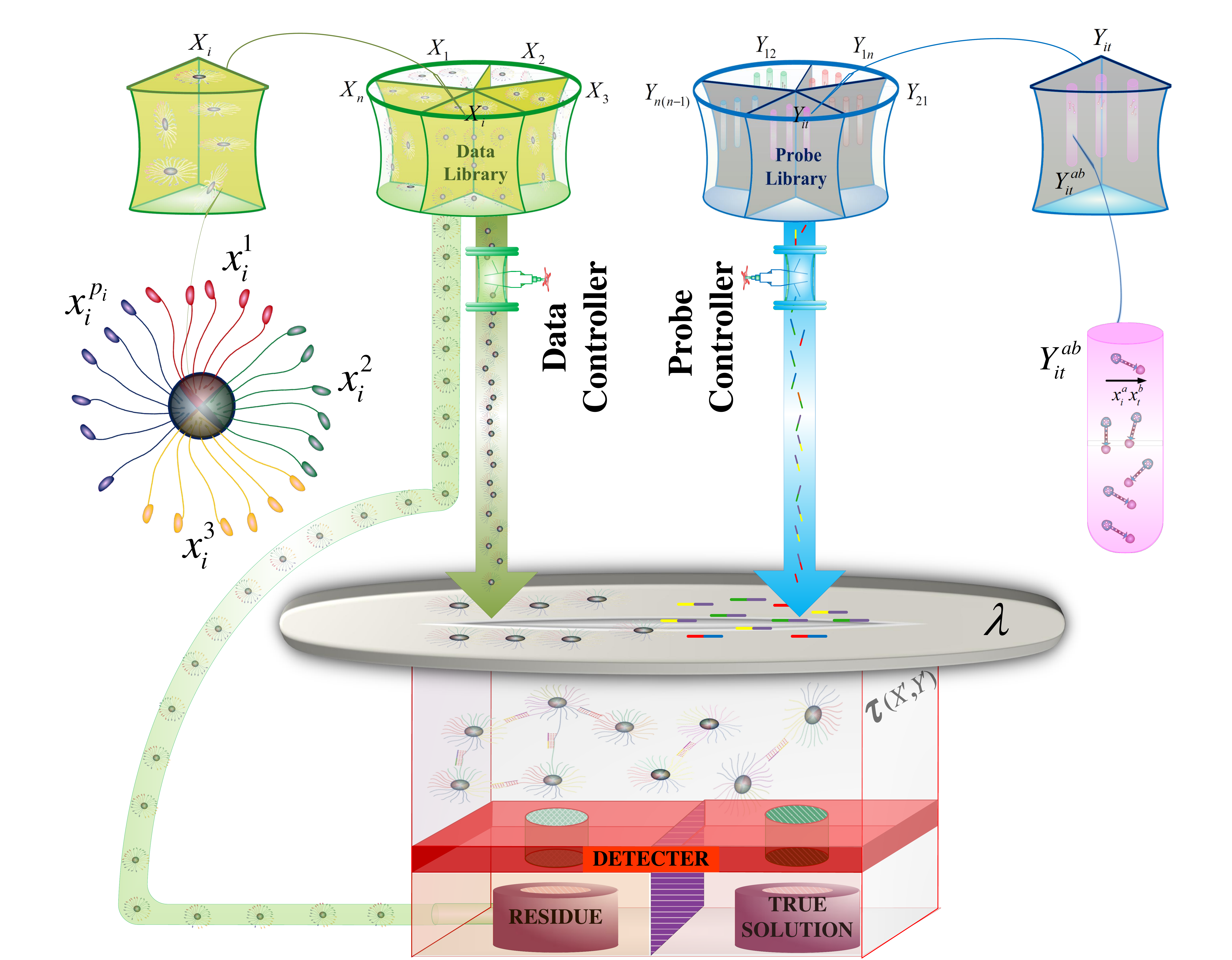}\\
 (b)
  \caption{ Two architectures of PMs. (a) Connective PM. (b) Transitive PM.}
    \label{figure8} 
\end{figure}

Based on previous discussions, we show two architectures of PM: the connective and transitive PMs, in Figure \ref{figure8}. To further demonstrate the computability of PM, we study the Hamilton cycle problem using the PM model in Subsection \ref{subsection3.10}. In Section \ref{section4}, we presents a \emph{nano-DNA PM model} to implement the connective PM, and show its efficacy in solving the graph coloring problem.

\subsection{Solving Hamilton cycle problem by Probe Machine}\label{subsection3.10}


A cycle of a graph $G$ is said to be a \emph{Hamilton cycle} if it contains all vertices of $G$. Such cycle was first considered by Hamilton\cite{Hamilton} in 1856.
A graph is \emph{hamiltonian} if it contains a Hamilton cycle. So far, no nontrivial necessary and sufficient condition for a graph to be hamiltonian has been known. Indeed, the problem of finding such a condition is one of the main unsolved problems in graph theory.
In this section, we establish a connective PM to solve this problem.

\begin{figure}[H]
\centering
\includegraphics[width=240pt]{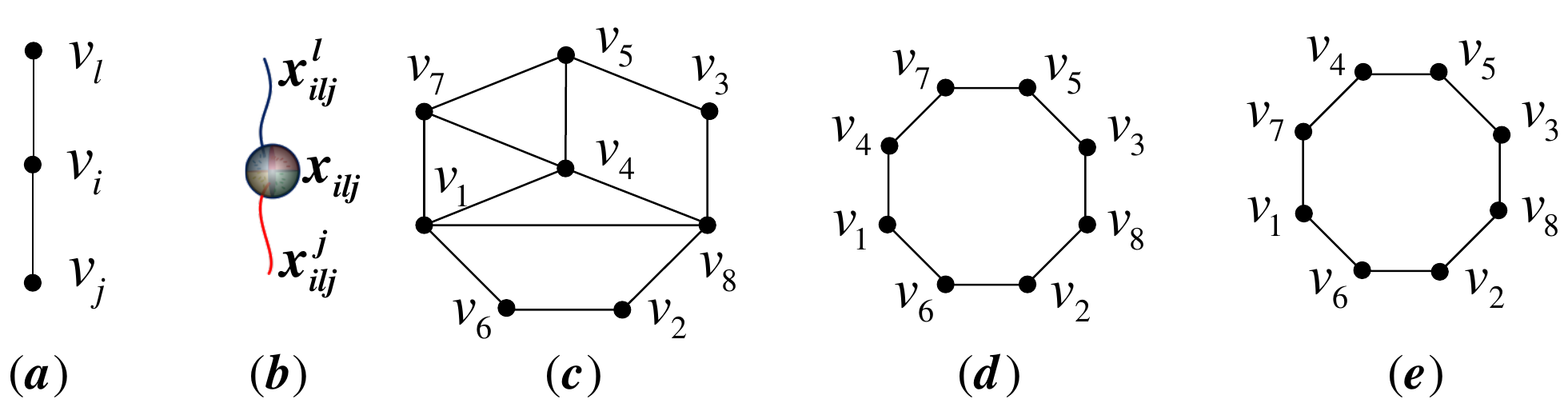}\\
\caption{(a) A 2-path with internal vertex $v_i$. (b) Data $x_{ilj}$ and its two data fibers $x_{ilj}^l$,$x_{ilj}^j$ in PM. (c) A graph with 8 vertices. (d) and (e) Two Hamilton cycles of the given graph. }
\label{figure9}
\end{figure}

Let $G$ be a simple undirected graph. We denote by $V(G)$ and $E(G)$ the sets of vertices and edges of $G$, respectively. For a vertex $v$, we use $\Gamma_G(v)$, or simply $\Gamma(v)$ when there is no scope for ambiguity, to denote the set of vertices adjacent to $v$ in $G$. Let $V(G)=\{v_1,v_2,\cdots,v_n\}$, $E(v_i)$ be the set of edges incident to $v_i$, and $E^2(v_i)$ be the set of all 2-paths with internal vertex $v_i$ (see Figure \ref{figure9} (a)), where an $\ell$-path is a path of length $\ell$.  Formally

$
E^2(v_i)=\Big{\{}v_lv_iv_j\triangleq x_{ilj} | v_l, v_j \in \Gamma(v_i);
$
\begin{equation}\label{eq.6}
 \hspace{2cm}\emph{i,j,l~are~mutually~ different}.\Big{\}}
\end{equation}

\noindent Based on $E^2(v_i)$, we construct the data library of the connective PM as follows.

\begin{equation}\label{eq.7}
X=\cup_{i=1}^{n} E^2(v_i)=\cup_{i=1}^{n} \{x_{ilj} | v_l, v_j \in \Gamma(v_i); i\neq j, l; l\neq j\}
\end{equation}

\noindent where $x_{ilj}$ has exactly two types of data fibers $x_{ilj}^{l}$  and  $x_{ilj}^{j}$ (see Figure \ref{figure9} (b)).

We now construct the probe library $Y$ based on data library $X$. Assume that $|V(G)|\geq 5$. We need to consider the following two cases.

\emph{Case} 1: $v_iv_t\not\in E(G)$. Let $x_{ilj}$ and $x_{tab}$ be two data in $X$. Then there exist a probe between them if and only if
\begin{equation}
\vert\{i,l,j\}\ \cap\{t,a,b\}\vert=\vert\{l,j\}\cap\{a,b\}\vert=1
\end{equation}

\noindent This equation indicates that $i\not \in \{t,a,b\}$, $t\not\in \{l,j\}$, and $\{l,j\}\cap \{a,b\}$ contains only one element.

\emph{Case} 2: $v_iv_t\in E(G)$. Then, there exists a probe for two data $x_{ilj}$ and $x_{tab}$ if and only if one of the following conditions holds.
\begin{enumerate}
\item[(1)] $\vert\{i,l,j\}\ \cap\{t,a,b\}\vert=\vert\{l,j\} \cap\{a,b\}\vert=1$
\item[(2)] $t\in\{l,j\},i\in\{a,b\}$, and $\{l,j\}\cap \{a,b\}=\emptyset$
\end{enumerate}

In fact, when solving the Hamilton cycle problem by PM, it is not necessary to use all of 2-paths of $G$, i.e. $\bigcup_{i=1}^{|V(G)|}E^2(v_i)$, as the data library. It suffices to consider the 2-paths with internal vertices in a vertex covering. Of course, a minimum covering is optimal.

By taking the graph in Figure \ref{figure9}(c) as an example, the following steps describe the procedure of solving Hamilton Cycle problem by the connective PM.

\emph{Step} $1$: Data library construction.

Evidently, $\{x_1,x_2,x_3,x_4,x_5\}$ is a minimum covering, and so the data library can be built by
\[X=E^2(v_1)\cup E^2(v_2)\cup E^2(v_3)\cup E^2(v_4)\cup E^2(v_5),\]
\noindent where $E^2(v_1)$ = $\{x_{174}, x_{178}, x_{176}, x_{148}, x_{146}, x_{186}\}$, $E^2(v_2)$ = $\{x_{268}\}$, $E^2(v_3)$ = $\{x_{358}\}$, $E^2(v_4)$ = $\{x_{458}, x_{451}, x_{457}, x_{481}, x_{487}, x_{417}\}$, and $E^2(v_5)$ = $\{x_{534}, x_{537}, x_{547}\}$. There are a total of 17 types of data in $X$, so we have
$34$ types of data fibers as follows.

\noindent $\mathfrak{J}(x_{174})=\{x^{7}_{174},x^{4}_{174}\}$, $\mathfrak{J}(x_{178})=\{x^{7}_{178},x^{8}_{178}\}$,  $\mathfrak{J}(x_{176})=\{x^{7}_{176},x^{6}_{176}\}$, $\mathfrak{J}(x_{148})=\{x^{4}_{148},x^{8}_{148}\}$, $\mathfrak{J}(x_{146})=\{x^{4}_{146},x^{6}_{146}\}$, $\mathfrak{J}(x_{186})=\{x^{8}_{186},x^{6}_{186}\}$, $\mathfrak{J}(x_{268})=\{x^{6}_{268},x^{8}_{268}\}$, $\mathfrak{J}(x_{358})=\{x^{5}_{358},x^{8}_{358}\}$, $\mathfrak{J}(x_{458})=\{x^{5}_{458},x^{8}_{458}\}$, $\mathfrak{J}(x_{451})=\{x^{5}_{451},x^{1}_{451}\}$, $\mathfrak{J}(x_{457})=\{x^{5}_{457},x^{7}_{457}\}$, $\mathfrak{J}(x_{481})=\{x^{1}_{481},x^{8}_{481}\}$, $\mathfrak{J}(x_{487})=\{x^{7}_{487},x^{8}_{487}\}$, $\mathfrak{J}(x_{417})=\{x^{7}_{417},x^{1}_{417}\}$,  $\mathfrak{J}(x_{534})=\{x^{3}_{534},x^{4}_{534}\}$, $\mathfrak{J}(x_{537})=\{x^{3}_{537},x^{7}_{537}\}$, $\mathfrak{J}(x_{547})=\{x^{4}_{547},x^{7}_{547}\}$.

\emph{Step} $2$: Probe library construction.

Based on the 34 types of data fibers, we construct the corresponding probe library according to the two rules mentioned above.

\noindent $Y_{12}=\{\overline{x^{8}_{178},x^{8}_{268}},\overline{x^{6}_{176},x^{8}_{268}},\overline{x^{8}_{148},x^{8}_{268}},\overline{x^{6}_{146},x^{6}_{268}} \}$,

\noindent $Y_{13}=\{\overline{x^{8}_{178},x^{8}_{358}},\overline{x^{8}_{148},x^{8}_{358}},\overline{x^{8}_{186},x^{8}_{358}}\}$,

 \noindent $Y_{14}=
 \{\overline{x^{7}_{178},x^{7}_{457}},\overline{x^{8}_{178},x^{8}_{458}},\overline{x^{7}_{176},x^{7}_{457}},\overline{x^{7}_{176},x^{7}_{487}},\\
\overline{x^{8}_{186},x^{8}_{458}},\overline{x^{8}_{186},x^{8}_{487}},\overline{x^{4}_{174},x^{1}_{451}},\overline{x^{4}_{174},x^{1}_{481}},\overline{x^{4}_{148},x^{1}_{451}},\\
 \overline{x^{4}_{148},x^{1}_{417}},
\overline{x^{4}_{146},x^{1}_{451}}, \overline{x^{4}_{146},x^{1}_{481}},\overline{x^{4}_{146},x^{1}_{417}} \}$,

  \noindent $Y_{15}=\{
 \overline{x^{7}_{174},x^{7}_{537}},\overline{x^{4}_{174},x^{4}_{534}},\overline{x^{7}_{178},x^{8}_{537}},\overline{x^{7}_{178},x^{7}_{547}},
 \overline{x^{7}_{176},x^{7}_{537}},\\ \overline{x^{7}_{176},x^{7}_{547}},\overline{x^{4}_{148},x^{4}_{534}},\overline{x^{4}_{148},x^{4}_{547}},
  \overline{x^{4}_{146},x^{4}_{534}},\overline{x^{4}_{146},x^{4}_{547}} \}$,

  \noindent $Y_{23}=\{
 \overline{x^{8}_{268},x^{8}_{358}} \}$,

 \noindent  $Y_{24}=\{
 \overline{x^{8}_{268},x^{8}_{458}},\overline{x^{8}_{268},x^{8}_{481}},\overline{x^{8}_{268},x^{8}_{487}} \}$,

 \noindent  $Y_{34}=\{
 \overline{x^{5}_{358},x^{5}_{451}},\overline{x^{5}_{358},x^{5}_{457}},\overline{x^{8}_{358},x^{8}_{481}},\overline{x^{8}_{358},x^{8}_{487}} \}$,

  \noindent $Y_{35}=\{
 \overline{x^{5}_{358},x^{3}_{534}},\overline{x^{5}_{358},x^{3}_{537}} \}$,

 \noindent  $Y_{45}=\{
 \overline{x^{7}_{487},x^{7}_{537}},\overline{x^{7}_{417},x^{7}_{537}},\overline{x^{5}_{458},x^{4}_{534}},\overline{x^{5}_{458},x^{4}_{547}},\overline{x^{5}_{451},x^{4}_{534}},\\ \overline{x^{5}_{451},x^{4}_{547}},\overline{x^{5}_{457},x^{4}_{534}} \}$.

\emph{Step} 3: Probe operations.
	
The data controller $\sigma_1$ takes an appropriate amount of data $x_{174}$, $x_{178}$, $x_{176}$, $x_{148}$,$x_{146}$ , $x_{186}$, $x_{268}$, $x_{358}$, $x_{458}$, $x_{451}$, $x_{457}$, $x_{481}$, $x_{487}$, $x_{417}$, $x_{534}$, $x_{537}$, $x_{547}$ from data library $X$, and adds them into the computing platform $\lambda$. Meanwhile, the probe controller $\sigma_2$ takes an appropriate amount of probes from probe sub-library $Y_{12}$, $Y_{13}$, $Y_{14}$, $Y_{15}$, $Y_{23}$, $Y_{24}$, $Y_{34}$, $Y_{45}$, and adds them into $\lambda$, and then the probe operation $\tau$ is executed to obtain the solutions to the problem.
	
\emph{Step} 4: True solution detection.
	
	The 4- or 5-aggregations are put into true solution storage by the detector. Others are put into the residue collector. Two Hamilton cycles, shown in Figure \ref{figure9}(d), \ref{figure9}(e), are obtained, and they are all of the Hamilton cycles of the graph shown in Figure \ref{figure9}(c).

\section{An implementation technique for the connective Probe Machine}\label{section4}
Probe Machine does exist in the natural world. For example, a biological nervous system is a typical connective PM. There are some neurons living with their organism for their lifetime, such as the neurons of cerebral cortex; meanwhile, there exist stem cells in some encephalic region. These stem cells can grow new neurons at the place where old neurons had died.  Of course, there also exist some neurons that cannot be replaced by new ones when they die.
Hence, one can see that nervous network is actually a dynamic network. It is well-known that information transmission among neurons is realized through
the synapses. There are totally two types of synapses: one is chemical synapses and the other is electrical synapses.
\begin{itemize}
  \item Chemical synapses have one-way information transfer mode, which uses chemical substance (neurotransmitter) as the communication medium.
  \item In contrast, the information transfer mode of electrical synapses is almost bidirectional, where the information is transmitted in the form of
electricity (electric signal).
\end{itemize}
Although there are sharp differences between these two types of synapses, information can only be transmitted under the effect of
action potential. Thus, we consider the action potential as the \emph{basic probe operation}.


Where and how to find appropriate materials to implement the PM is the key to develop a PM-based computer. 
This section presents a model called \emph{nano-DNA PM model} to implement the connective PM. In a nano-DNA PM model, the data is made by the complexus of nanoparticle and DNA molecule, and the probe is made by DNA molecule.
By applying the current nanotechnology and biotechnology, especially the detection technology, it is still difficult to develop a nano-DNA PM model to solve large-scale practical problems \cite{17}. Nevertheless, with the development of detection technology, hopefully a nano-DNA probe computer can be put into practice.

As for the implementation of the transitive PM, we have a conjecture: the data can be complex, the information in data fiber can be neurotransmitter (such as acetylcholine), and the probe can be analogous to the action potential in the biological nervous system. This research will be elaborated in another paper.

In what follows, we introduce the main components of a nano-DNA PM model.

\emph{Data library construction}. For constructing the data in data library, the data body is made by nanoparticle and the data fiber is made by DNA chain (see Figure \ref{figure2} (c) and \ref{figure2}(d)). We visually refer to this kind of data as \emph{starlets}. We now present the detailed process of producing a starlet $x$.
First, divide the nanoparticle equally into $p_{i}$ connected regions, where $p_{i}$ is the number of different types of data fibers in $x$; second, connect the same type of data fibers (DNA sequences) as many as possible in each region. The method of connecting DNA sequences is described in \cite{19} with details.

\emph{Probe library construction}: Let $x_{i}^{a}$ and $x_{t}^{b}$ be the fibers of data $x_{i}$ and $x_{t}$, respectively. The probe of these two data fibers,
denoted by $\overline{x_{i}^{a} x_{t}^{b}}$,  is composed by the complementary DNA chain that consists of half DNA chain of $x_{i}^{a}$
and half DNA chain of $x_{t}^{b}$, respectively (see Figure \ref{figure3}).

\emph{Detector}. It is challenging to design a practical detector.
Currently, in biochemical tests, detections are mainly processed by $PCR$ amplifier or
electronic speculum which are not sufficiently-reliable techniques. Once the detection technology becomes mature, the nano-DNA probe machine can be put into practice immediately.

\begin{figure}[hpb]
  \centering
   \includegraphics[width=140pt]{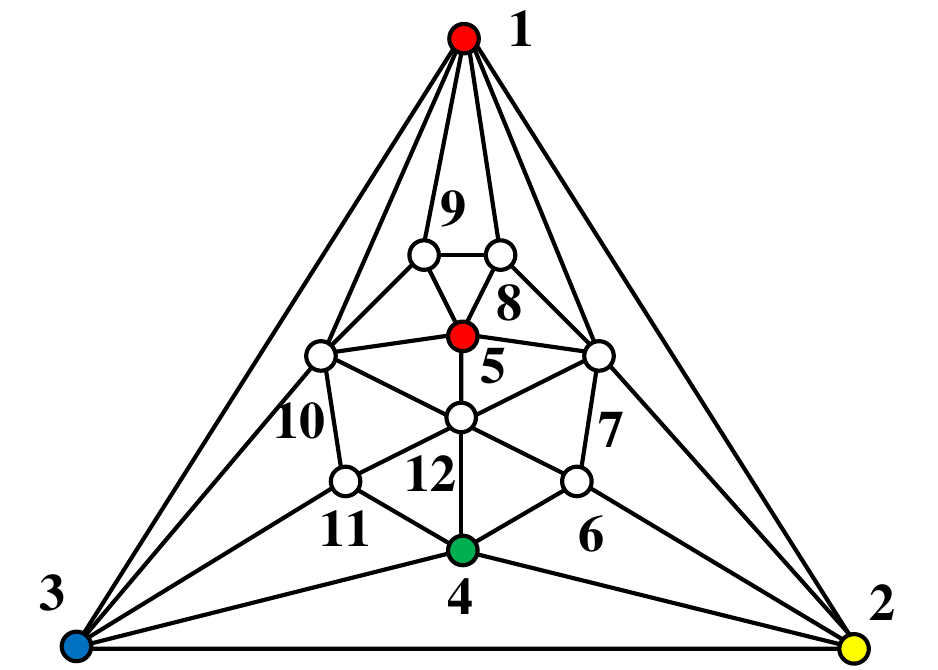}
  \caption{A 4-colorable maximal planar graph $G_{12}$ on 12 vertices and $V^{\prime\prime}(G_{12})$. }
    \label{figure10} 

\end{figure}

As an illustration, consider a 4-coloring problem of the graph $G_{10}$ shown in Figure \ref{figure10}.
A (proper) $k$-coloring of a graph $G$ is an assignment of $k$ colors to $V(G)$, the set of vertices of $G$, such that no two adjacent vertices can be assigned with the same color.
It can be viewed as a partition $\{V_1,V_2,\cdots,V_k\}$ of $V(G)$, where $V_i$ is an independent set and is called a \emph{color class} of the coloring.
For a graph $G$, we denote by $V^{\prime\prime}(G)$ the set of vertices, each of which  belongs to an unchanged color class for all 4-colorings of $G$ (up to color permutation). One can readily check that  $V^{\prime\prime}(G_{12})=\{1,2,3,4,5\}$. The basic rule for designing probe is that any two adjacent vertices have no same colored probes.

Based on the 4-coloring of $V^{\prime\prime}(G_{12})$, we build Table \ref{table1}. In this table, the numbers $1,2,\cdots,12$ in the first row represent the labels of  vertices of $G_{12}$. Each vertex can be colored by one of the colors given in its column,
where $r$, $y$, $b$, $g$ denote red, yellow, blue, and green, respectively and their subscripts denote vertex labels.

\begin{table}%
\renewcommand{\arraystretch}{1.3}
\caption{21 possible colors assigned to the vertices of $G_{12}$\label{table1}}{%
\begin{tabular}{|c|c|c|c|c|c|c|c|c|c|c|c|}
    \hline
    1 & 2 & 3 & 4 & 5 & 6 & 7 & 8 & 9 & 10 & 11 & 12 \\
    \hline
    $r_{1}$ & & & & $r_{5}$ & $r_{6}$ & & & & & $r_{11}$ & \\
    \hline
     & $y_{2}$ & & & & & & $y_{8}$ & $y_{9}$ & $y_{10}$ & $y_{11}$ & $y_{12}$\\
     \hline
     & & $b_{3}$ & & & $b_{6}$ & $b_{7}$ & $b_{8}$ & $b_{9}$ & & & $b_{12}$ \\
     \hline
     & & & $g_{4}$ & & & $g_{7}$ & $g_{8}$ & $g_{9}$ & $g_{10}$ & & \\
     \hline
\end{tabular}}
\end{table}%

The data library and probe library are constructed as follows.

1. \emph{Data library}

Based on Table \ref{table1}, we construct the data library by 12 types of data, denoted by $x_1, x_2, \cdots, x_{12}$, respectively. For each type of data $x_i$, its body is labeled as $i$, and its fibers are all of the possible colors of the vertex $i$. The sets of data fibers of these 12 types of data are shown in the columns of Table \ref{table1}, i.e.

 \begin{center}
     ${\Im _1} = \{ {r_1}\}$,  ${\Im _2} = \{ {y_2}\}$, ${\Im _3} = \{ {b_3}\}$,  ${\Im _4} = \{ {g_4}\}$,
    ${\Im _5} = \{ {r_5}\}$,  ${\Im _6} = \{ {r_6, b_6}\}$,  ${\Im _7} = \{ {b_7, g_7}\}$,  ${\Im _8} = \{ {y_1, b_8, g_8}\}$,
    ${\Im _9} = \{ {y_9, b_9, g_9}\}$,${\Im _{10}} = \{ {y_{10}, g_{10}}\}$, ${\Im _{11}} = \{ {r_{11}, y_{11}}\}$,  ${\Im _{12}} = \{ {y_{12}, b_{12}}\}$.
 \end{center}

2. \emph{Probe Library}

    \begin{table*}[ht]
    \renewcommand{\arraystretch}{1.3}
    \caption{The sets of 4-color probes corresponding to the edges of $G$, totally 30 sub-probe libraries, and 73 types of probes\label{table2}}{
    \centering
   ~~~~~~~~~~~~~~~~~~~~~~ \begin{tabular}{|c|c|c|c|c|c|c|c|c|c|c|}
    \hline $x_1x_2$ &  $x_1x_3$ &  $x_1x_7$ &  $x_1x_8$ &   $x_1x_9$ &  $x_1x_{10}$ &  $x_2x_3$ &  $x_2x_4$ &   $x_2x_6$ &  $x_2x_7$ & $x_3x_4$ \\ \hline $\overline{r_1y_2}$ & & & $\overline{r_1y_8}$&  $\overline{r_1y_9}$ & $\overline{r_1y_{10}}$ & & & $\overline{r_2y_6}$ & & \\ \hline & $\overline{r_1b_3}$& $\overline{r_1b_7}$&  $\overline{r_1b_8}$ & $\overline{r_1b_{9}}$ & & $\overline{y_2b_3}$& &  $\overline{y_2b_6}$ & $\overline{y_2b_7}$ & \\ \hline & &  $\overline{r_1g_7}$&  $\overline{r_1g_8}$ & $\overline{r_1g_{9}}$ & $\overline{r_1g_{10}}$ && $\overline{y_2g_4}$ &  & $\overline{y_2g_7}$ & $\overline{b_3g_{4}}$ \\ \hline $x_3x_{10}$ &  $x_3x_{11}$ &  $x_4x_6$ &  $x_4x_{11}$ &   $x_4x_{12}$ &  $x_5x_7$ &  $x_5x_8$ &  $x_5x_9$ &   $x_5x_{10}$ &  $x_5x_{12}$ & $x_6x_7$ \\ \hline $\overline{b_3y_{10}}$ &  $\overline{b_3y_{11}}$ &   &  $\overline{g_4r_{11}}$ &   $\overline{g_4b_{12}}$ &  $\overline{r_5b_7}$ &  $\overline{r_5y_8}$ &  $\overline{r_5y_9}$ &   $\overline{r_5y_{10}}$ &  $\overline{r_5y_{12}}$ & $\overline{r_6b_7}$ \\ \hline $\overline{b_3g_{10}}$ &  $\overline{b_3y_{11}}$ &  $\overline{g_4r_6}$ &  $\overline{g_4y_{11}}$ &   $\overline{g_4y_{12}}$ &   &  $\overline{r_5b_8}$ &  $\overline{r_5b_9}$ &    &  $\overline{r_5b_{12}}$ & $\overline{r_6g_7}$ \\ \hline &    &  $\overline{g_4b_6}$ &    &     &  $\overline{r_5g_7}$ &  $\overline{r_5g_8}$ &  $\overline{r_5g_9}$ &   $\overline{r_5g_{10}}$ &  & $\overline{b_6g_7}$ \\ \hline $x_6x_{12}$ &  $x_7x_{8}$ &   &  $x_7x_{12}$ &   $x_8x_{9}$ &   &  $x_9x_{10}$ &   &   $x_{10}x_{11}$ &  $x_{10}x_{12}$ & $x_{11}x_{12}$ \\ \hline $\overline{r_6y_{12}}$ &  $\overline{b_7y_{8}}$ &  $\overline{g_7b_{8}}$  &  $\overline{b_7y_{12}}$ &   $\overline{y_8b_{9}}$ & $\overline{b_8g_{9}}$  &  $\overline{y_9g_{10}}$ &  $\overline{g_9y_{10}}$   &   $\overline{y_{10}r_{11}}$ &  $\overline{y_{10}b_{12}}$ & $\overline{r_{11}y_{12}}$ \\ \hline $\overline{y_6b_{12}}$ &  $\overline{b_7g_{8}}$ &   &  $\overline{g_7y_{12}}$ &   $\overline{y_8g_{9}}$ &   $\overline{g_9y_{8}}$  &   $\overline{b_{9}y_{10}}$ & &  $\overline{g_{10}r_{11}}$& $\overline{g_{10}y_{12}}$ & $\overline{r_{11}b_{12}}$ \\ \hline $\overline{b_6y_{12}}$ &  $\overline{g_7y_{8}}$ &   &  $\overline{g_7b_{12}}$ &   $\overline{b_8y_{9}}$ &  $\overline{g_8b_{9}}$  &  $\overline{b_9g_{10}}$ &   &   $\overline{g_{10}y_{11}}$ &  $\overline{g_{10}b_{12}}$ & $\overline{y_{11}b_{12}}$ \\ \hline \end{tabular}}
     \end{table*}

      In what follows, we denote by $x_i$, for $i=1,2, \cdots, 12$, the vertex $i$ in the graph shown in Figure \ref{figure10}. We design the probes according to the edges of $G$. For example, for the edge $x_4x_{12}$, since $x_{12}$ has two available colors,
 there should have two probes $\overline{g_4y_{12}}$  and $\overline{g_4b_{12}}$.

 Each column in Table \ref{table2} corresponding to $x_ix_t$ includes all of probes of probe sub-library $Y_{it}$ for $i,t=1,2,\cdots, 12$ and $i\neq t$. There are a total of 30 sub-probe libraries and 73 types of probes.

    The detailed procedure of probe operations is presented as follows.

    \emph{Step} 1. First, make 12 types of nanoparticles (2.5nm) as the data bodies. Second, encode the DNA sequences corresponding to the 21 types of data fibers and synthesize the DNA strands. Finally, embed the DNA strands (data fibers) into their corresponding nanoparticles (data bodies).

      \emph{Step} 2. Based on Step 1, construct the probe library consisting of 73 types of probes.

     \emph{Step} 3. Take appropriate amount of data (12 types) from data library and (73 types) probes from probe library, and then add them into the computing platform. Many data aggregations can be formed by specific hybridization among DNA molecules. Note that at present no technology can implement a computing platform with the proposed three properties. However, a nano-DNA PM model can still form the aggregations containing all of the true solutions as long as the numbers of the data and probes are sufficient.

     \emph{Step} 4. Detect all of the 12-aggregations isomorphic to the graph shown in Figure \ref{figure10}. Note that the current TEM (electron microscope) can only detect the aggregations with a small rather than large order, which is where the main difficulty lies for taking PM into practice.

     \begin{figure}[H]                                                                                                                                                                                                                                       \centering
      \includegraphics[width=260pt]{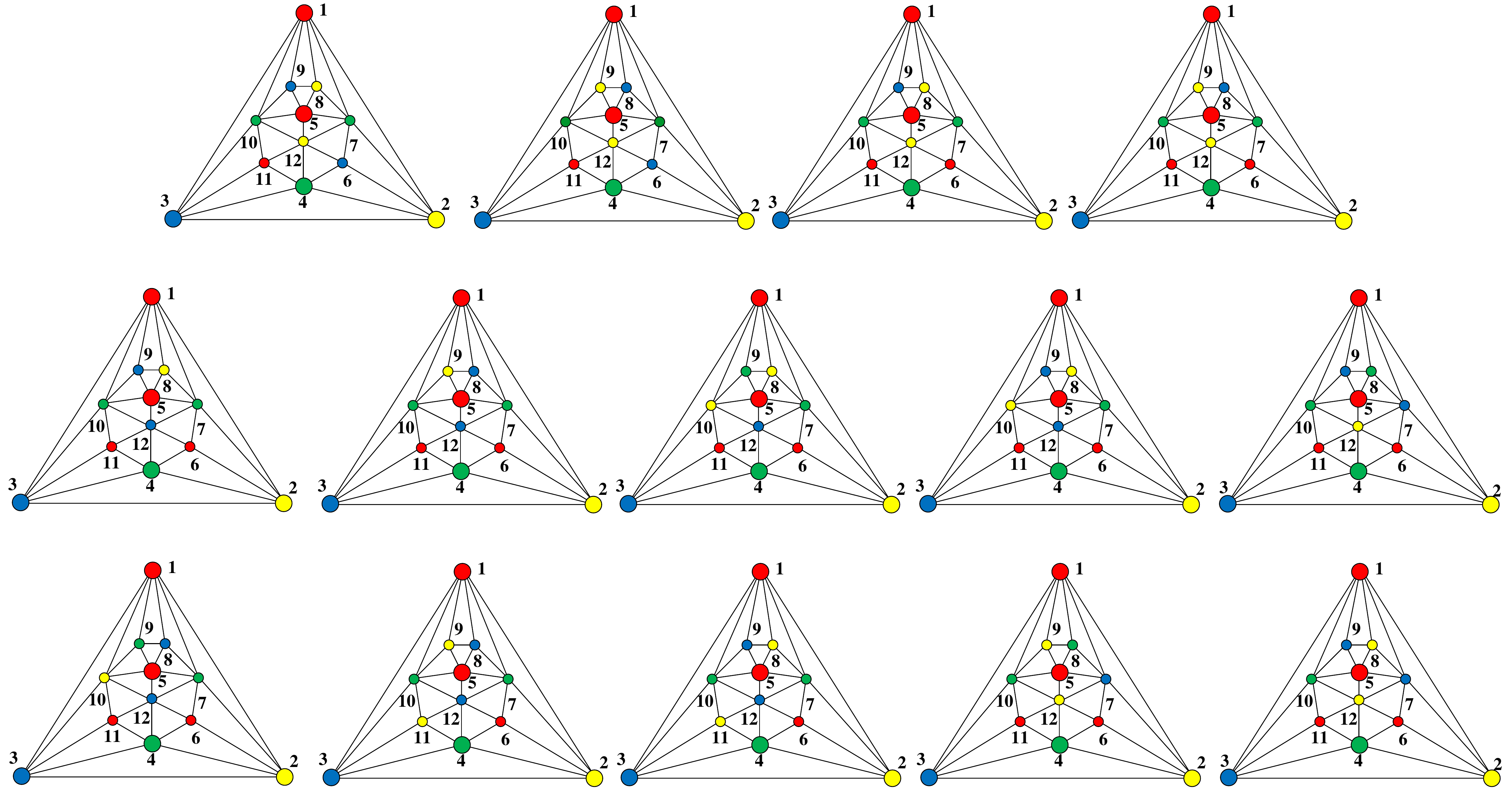}
     \caption{All 4-colorings of the graph.}
      \label{figure11}
     \end{figure}

     Theoretically, we can find all of the 14 true solutions---i.e. the 14 different colorings (shown in Figure \ref{figure11})---through only one probe operation.

\section{Performance Analysis of Probe Machine}

Compared to existing computing models, PM model has the following advantages regarding the performance in computation.

(1) TM is a special case of PM, and any known NP-complete problem (based on TM) can be solved in one or several probe operations under the PM model.

The fundamental parallelism in computation of PM significantly reduces the computational complexity of addressing NP-complete problems to one or several probe operations. Suppose $X^{\prime}\subseteq X$ , $Y^{\prime}\subseteq\tau(X^{\prime})$, $\Theta=\tau(X^{\prime},Y^{\prime})$. Then $\Theta$ may contain either only one true solution or thousands of true solutions.
The topological structure of each true solution is isomorphic to $G^{(X',Y')}$, although the weight of each vertex may differ. In fact, the weight of a vertex corresponds to data fibers of this vertex. In the process of each probe operation, the number of basic probe operations is equal to the summation of numbers of edges over all of solution graphs. The computational parallelism of PM can help address each of those known NP-complete problems by taking only one or several probe operations, including the Hamilton cycle problem, the graph coloring problem (see Subsection \ref{subsection3.10} and Section \ref{section4}).  In 1971, Cook et al. proved that the complexity of NP-complete problems are mutually equivalent in polynomial time under TM \cite{21}. Each operation in TM moves its read-write head to left or to right by one grid, erases the original data, and writes down the result in the grid. The topological structure of TM in each operation is a path of length one, which corresponds to an edge of a graph among all solution graphs.

(2) The feasibility of implementing connective and transitive PMs.

The intuition behind the proposal of the transitive probe is based on the  observation that the information transmission between neurons in a natural biosystem is through synapse under the action potential. Thus, we take the \emph{action potential} as the transitive probe. Although the transitive PM simulates the information transmission in a biosystem, it has two distinguishing properties. First, the positions of two arbitrary neurons in a biosystem are fixed, while the positions of data fibers in the transitive PM  are not. During a transitive operation, information transmission can only occur after PM finds two data fibers in the computing platform. Second, the connections among neurons in biosystem are very sparse (in human brain, there are about $10^{12}$ neurons but only $10^{15}$$\thicksim$$10^{16}$ synapses), while the connections among data in PM are dense---even any pair of data in PM is potential.

(3) The information processing capability of PM dramatically increases with the size of data library getting large.

 Generally, the information processing capability of PM in one operation is $2^{q}$, where $q$ is the number of all edges over solution graphs isomorphic to $G^{(X',Y')}$ in $\Theta$. For example, when the size of data library is 50 and each pair of data has potential for information transmission, the processing capability of PM will reach to $2^{25\times49}$=$2^{1225}$, which is sufficient to break any public key crypto-system.

\section{Conclusions and Future Work}
\label{conclusion}
TM is a computing model with a linear data placement mode, and only adjacently-placed data can be processed sequentially by individual operations. In contrast, PM proposed in this paper is a fully-parallel computing model with a data placement that is free of constrains in the space, and any pair of data can be viewed as adjacent, and can be processed directly. Therefore, PM has the powerful computation capability of processing numerous pairs of data simultaneously. It only requires one or several probe operations to solve some hard problems, which TM cannot solve in polynomial time.

PM consists of nine components, among which the data library and probe library are the core components. The probes are divided into two types: connective and transitive probes. The connective operation does not consider the direction of information processing between data fibers, while the transitive operation does. Furthermore, we conjecture that probes  may only have these two types. In the demonstrations of solving the Hamilton and graph coloring problems, the basic operations of PM used are connective operations, while the information processing in  biological neurons is transitive. Do other types of probes  beyond these two exist? Is there any mixed PM that contains both connective and transitive probes? These interesting questions are worthy of exploring in our future work.

PM proposed in this paper is a mathematical model, whose computation capability and effectiveness exceed those of TM. We believe PM-based computers will contribute to the development of human civilization and society. Now a practical problem is: What material can be leveraged to manufacture computers based on PM? Although we mentioned a nano-DNA PM model, it is far from a practical implementation at present. The transitive PM model holds the promise of being implemented in future~\cite{18,19}, and we believe that its capability may be more powerful than human thought in certain aspects.

\ifCLASSOPTIONcompsoc
  \section*{Acknowledgments}
\else
  \section*{Acknowledgment}
\fi

This work is supported by 973 projects (2013CB32960, 2013CB329602);
 National Natural Science Foundation of China Equipment (61127005); National Natural Science Foundation of China under grant 60974112, 30970960.

\ifCLASSOPTIONcaptionsoff
  \newpage
\fi

\end{document}